\documentclass[twocolumn,showpacs,amsmath,amssymb,prl,superscriptaddress]{revtex4-1}
\pdfoutput=1
\usepackage{graphicx}
\usepackage{dcolumn}
\usepackage{bm}
\usepackage{pdfpages}%
\begin{document}

\title{Bond-order and the Role of Ligand States in Stripe-Modulated IrTe$_2$}

\author{K. Takubo}
\email{ktakubo@issp.u-tokyo.ac.jp}
\affiliation{Department of Physics {\rm {\&}} Astronomy, University of British Columbia, Vancouver, British Columbia V6T\,1Z1, Canada}
\affiliation{Quantum Matter Institute, University of British Columbia, Vancouver, British Columbia V6T 1Z4, Canada}
\affiliation{Max Planck Institute for Solid State Research, Heisenbergstrasse 1, D-70569 Stuttgart, Germany}
\author{R. Comin}
\affiliation{Department of Physics {\rm {\&}} Astronomy, University of British Columbia, Vancouver, British Columbia V6T\,1Z1, Canada}
\affiliation{Quantum Matter Institute, University of British Columbia, Vancouver, British Columbia V6T 1Z4, Canada}
\author{D. Ootsuki}
\affiliation{Department of Physics {\rm {\&}} Department of Complexity Science and Engineering, University of Tokyo, 5-1-5 Kashiwanoha, Chiba 277-8581, Japan}
\author{T. Mizokawa}
\email{mizokawa@k.u-tokyo.ac.jp}
\affiliation{Department of Physics {\rm {\&}} Department of Complexity Science and Engineering, University of Tokyo, 5-1-5 Kashiwanoha, Chiba 277-8581, Japan}
\author{H. Wadati}
\affiliation{Institute for Solid State Physics, University of Tokyo, Kashiwanoha 5-1-5, Chiba 277-8581, Japan}
\author{{\mbox{Y. Takahashi}}}
\affiliation{Department of Physics, University of Tokyo, Hongo, Tokyo 113-0033, Japan}
\author{G. Shibata}
\affiliation{Department of Physics, University of Tokyo, Hongo, Tokyo 113-0033, Japan}
\author{A. Fujimori}
\affiliation{Department of Physics, University of Tokyo, Hongo, Tokyo 113-0033, Japan}
\author{R. Sutarto}
\affiliation{Canadian Light Source, University of Saskatchewan, Saskatoon, Saskatchewan S7N 2V3, Canada} 
\author{F. He}
\affiliation{Canadian Light Source, University of Saskatchewan, Saskatoon, Saskatchewan S7N 2V3, Canada}
\author{S. Pyon}
\affiliation{Department of Physics, Okayama University, Okayama 700-8530, Japan}
\author{K. Kudo}
\affiliation{Department of Physics, Okayama University, Okayama 700-8530, Japan}
\author{\mbox{M. Nohara}}
\affiliation{Department of Physics, Okayama University, Okayama 700-8530, Japan} 
\author{G. Levy}
\affiliation{Department of Physics {\rm {\&}} Astronomy, University of British Columbia, Vancouver, British Columbia V6T\,1Z1, Canada}
\affiliation{Quantum Matter Institute, University of British Columbia, Vancouver, British Columbia V6T 1Z4, Canada}
\author{I. S. Elfimov}
\affiliation{Department of Physics {\rm {\&}} Astronomy, University of British Columbia, Vancouver, British Columbia V6T\,1Z1, Canada}
\affiliation{Quantum Matter Institute, University of British Columbia, Vancouver, British Columbia V6T 1Z4, Canada}
\author{G. A. Sawatzky}
\affiliation{Department of Physics {\rm {\&}} Astronomy, University of British Columbia, Vancouver, British Columbia V6T\,1Z1, Canada}
\affiliation{Quantum Matter Institute, University of British Columbia, Vancouver, British Columbia V6T 1Z4, Canada}
\author{A. Damascelli}
\email{damascelli@physics.ubc.ca}
\affiliation{Department of Physics {\rm {\&}} Astronomy, University of British Columbia, Vancouver, British Columbia V6T\,1Z1, Canada}
\affiliation{Quantum Matter Institute, University of British Columbia, Vancouver, British Columbia V6T 1Z4, Canada}

\date{\today}

\begin{abstract}
The coupled electronic-structural modulations of the ligand states in IrTe$_2$ have been studied by x-ray absorption spectroscopy (XAS) and resonant elastic x-ray scattering (REXS). Distinctive pre-edge structures are observed at the Te-$M_{4,5}$ (3$d$ $\rightarrow$ 5$p$) absorption edge, indicating the presence of a Te 5$p$\,-\,Ir 5$d$ covalent state near the Fermi level. An enhancement of the REXS signal near the Te 3$d$ $\rightarrow$ 5$p$ resonance at the $Q\!=\!(1/5,0,-1/5)$ superlattice reflection is observed below the structural transition temperature $T_s\sim 280$\,K. The analysis of the energy-dependent REXS lineshape reveals the key role played by the spatial modulation of the covalent Te 5$p$\,--Ir 5$d$ bond-density in driving the stripe-like order in IrTe$_2$, and uncovers its coupling with the charge and/or orbital order at the Ir sites. The similarity between these findings and the charge-ordering phenomenology observed in the high-T$_c$ superconducting cuprates suggests that the iridates may harbor similar exotic phases.
\end{abstract}

\maketitle

Transition-metal compounds exhibit surprisingly rich electronic and magnetic properties due to the partially filled $d$ orbitals. The fundamental properties of the electronic structure of transition-metal compounds can be described within the Zaanen-Sawatzky-Allen (ZSA) scheme. This differentiates between the Mott-Hubbard regime ($U < \Delta$) and the charge-transfer regime ($\Delta < U$), depending on the relative balance of the on-site Coulomb interaction $U$ between the $d$ electrons and the charge-transfer energy $\Delta$ between the ligand states and the transition-metal $d$ states \cite{ZSA}. When $\Delta$ approaches zero, the ligand states are almost degenerate in energy with the transition-metal $d$ levels. As a result, the ligand states may participate in those spin, charge, and/or orbital ordering phenomena that are peculiar to the correlated nature of the $d$-orbitals. As an example of such phenomenology, ordering of the oxygen 2$p$ holes is realized in the stripe-ordered phase of layered cuprates \cite{Abbamonte05,Fink09,Fink11,Hawthorn11,Achkar13}, or in the ladder-type Cu oxides \cite{Abbamonte04}.

Very recently, a first-order structural transition was discovered in the 5$d$ transition-metal chalcogenide IrTe$_2$ at $T_s\!\sim\!280$\,K. This attracted great interest due to the concomitant discovery of superconductivity in the Pt- and Pd-substituted or intercalated compounds \cite{Pyon12,Yang12}. Clarifying the origin of the structural phase transition might be a critical step towards the understanding of  superconductivity itself; however, to date several mechanisms have been debated, with a universal consensus still lacking.
\begin{figure*}[t!]
\includegraphics[width=1\linewidth]{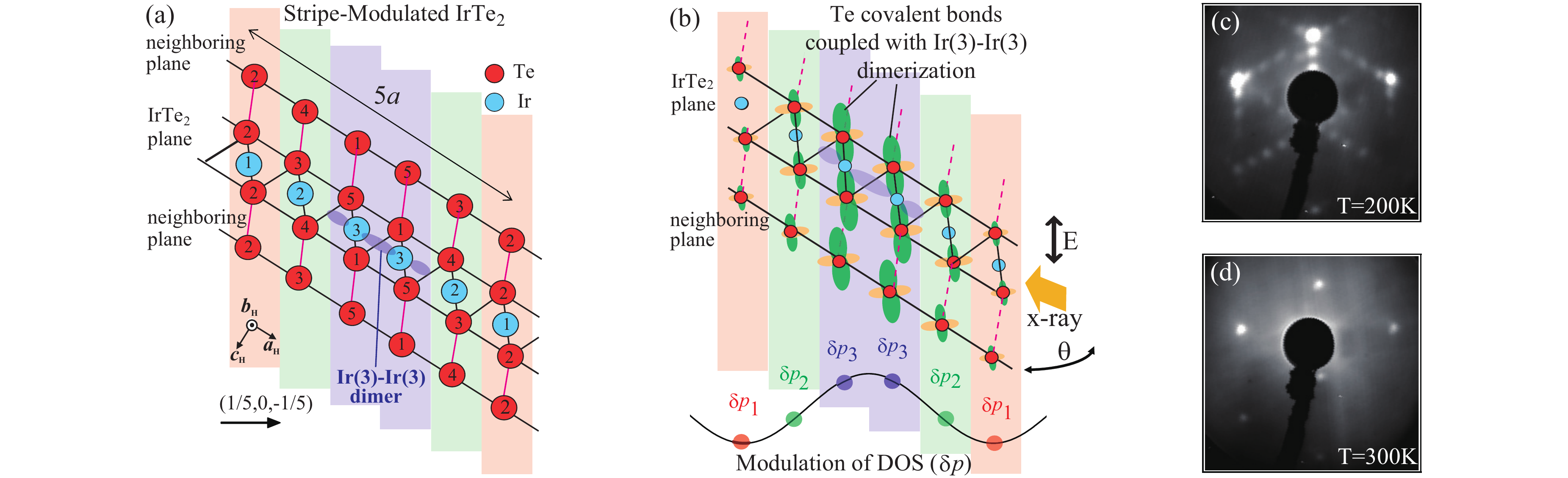}%
\caption{(color online). (a) IrTe$_2$ superstructure modulation with wavevector $Q\!=\!(1/5,0,-1/5)$, as expressed in reciprocal lattice units in tetragonal notation. Numeric labels denote the inequivalent Ir and Te sites. The modulation of the density-of-states (DOS) -- as estimated from dynamical mean-field theory (DMFT) \cite{Pascut13} and highlighting the Ir(3)-Ir(3) dimerization -- is shown at the bottom as well as above with correspondingly colored shading. (b) Illustration of the covalent bonds between the hybridized Te and Ir sites: the orbital size denotes the covalent character. In virtue of the experimental geometry (see text), REXS is sensitive to these covalent bonds. (c,d) LEED pattern measured on IrTe$_2$ at 200 and 300\,K, with 80\,eV electrons. 
}
\end{figure*}
The phase transition is accompanied by the emergence of a superstructure lattice modulation in electron diffraction \cite{Yang12} -- with wavevector $Q\!=\!(1/5,0,-1/5)$ as expressed in reciprocal lattice units in tetragonal notation -- which is here illustrated in Fig.\,1. The main elements are the Ir-Ir dimerization along the $a$ axis with period $5a$, and the consequent distortion of the triangular Ir sublattice in the $a-b$ plane, conflating to an overall trigonal-to-triclinic symmetry reduction. The Ir-Ir dimerization likely stabilizes a unique stripe-like order -- with stripes running along the $b$ axis -- as indicated by x-ray diffraction \cite{Pascut13,Toriyama13} and extended x-ray absorption fine structure \cite{Joseph13} studies. Such superstructure can be explained by the emergence of a charge-density wave (CDW) driven by perfect or partial nesting of the multi-band Fermi surface \cite{Yang12}. Since in IrTe$_2$ the formal valence of Ir is +4, the Ir 5$d$ electrons with $t_{2g}^5$ configuration are the closest to the chemical potential, and are thus expected to play a central role in a CDW. However, a photoemission study has shown that the charge-transfer energy $\Delta$ in IrTe$_2$ is close to zero, and that the Te 5$p$ states are also important for the low energy physics \cite{Ootsuki12}. As further emphasized by recent studies \cite{Fang13,Oh13,Qian13}, the Te 5$p$ states might possibly be even more important than the Ir 5$d$ states in the CDW phase transition of IrTe$_2$.

To resolve the controversy on the microscopic origin of the phase transition, the contribution of the Te 5$p$ states to the superstructure formation must be experimentally quantified. In this context, resonant elastic x-ray scattering (REXS) experiments at the Te 3$d$ $\rightarrow$ 5$p$ resonance represent the most effective method to directly probe the spatial ordering of the Te 5$p$ states. Here we use REXS on IrTe$_2$ to reveal a modulation of the Te 5$p$-Ir 5$d$ covalent-bond-state with the same wavevector $Q\!=\!(1/5,0,-1/5)$ as observed for the structural transition. This covalent-bond modulation is further coupled with the 5$d$ orbital states at the Ir sites, and is thus ultimately responsible for the stripe-like ordering formation in IrTe$_2$.

REXS and x-ray absorption spectroscopy (XAS) measurements were performed at the REIXS beamline of the Canadian Light Source \cite{Hawthorn_RevSci}. Single crystals of IrTe$_2$ were prepared using a self-flux method \cite{Fang13,Pyon13}, and then cleaved in situ to minimize surface contamination effects. For the REXS measurements, the incident light was polarized along the (1,0,-1) direction [Fig.\,1(b)]. XAS was used to determine the photoabsorption coefficient $\mu(\omega)$, which is proportional to the imaginary part of the form factor, $\mu(\omega) \!\propto\! \mathrm{Im} \{f_j(\hbar\omega)\}$. Low-energy electron diffraction (LEED) measurements were performed at UBC, with a SPECS ErLEED 100 set-up and an electron energy of 80\,eV, at 200 and 300\,K.

The IrTe$_2$ XAS spectra around the Te-$M_{4,5}$ edges (corresponding to the creation of a Te-3$d$ core hole) are plotted in Fig.\,2(a). Distinct pre-edge (labeled as $M_{4,5}$) and main edge structures can be clearly observed. While the final state of the larger main edge is the $s$-$d$-$f$ hybrid band (of 6$s$, 4$f$, and 5$d$ character), the final state for the pre-edges is the Te-5$p$ manifold. In light of previous experimental studies of these absorption channels \cite{Jiang04,Song08,Telesca11}, the pre-edge peak structure may be more precisely ascribed to transitions into Te-Ir covalent states.
\begin{figure*}[t!]
\includegraphics[width=1\linewidth]{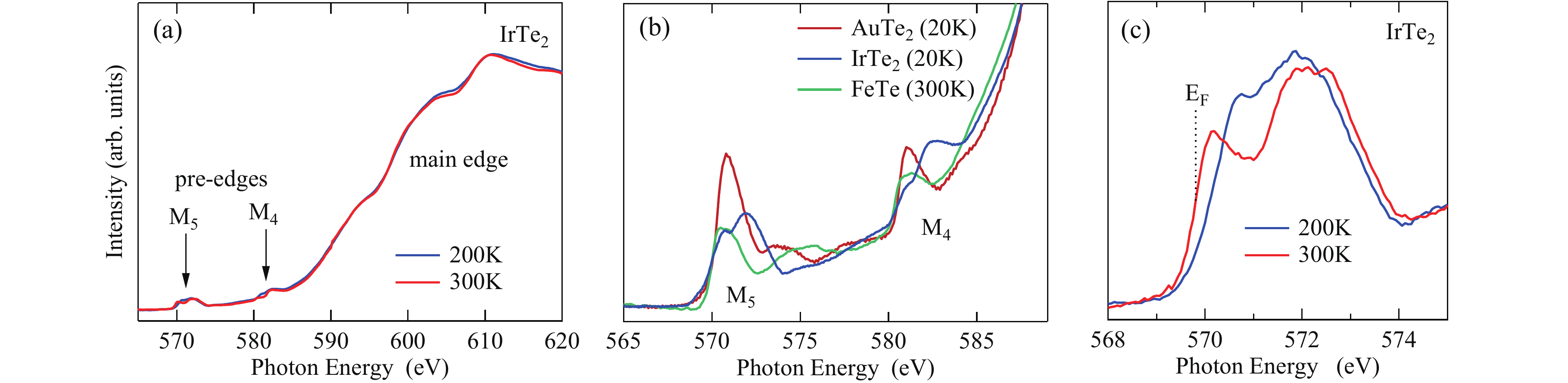}%
\caption{(color online). (a) IrTe$_2$ XAS spectra measured at the Te-$M$ absorption edge at 200 and 300\,K; the Te-$M_{4,5}$ pre-edge features are indicated. (b) Pre-edge XAS spectra from FeTe, IrTe$_2$, and AuTe$_2$. (c) IrTe$_2$ Te-$M_{5}$ XAS spectra at 200 and 300\,K.
}
\end{figure*}
For a more conclusive assignment, in Fig.\,2(b) we compare the pre-edge region for FeTe, IrTe$_2$, and AuTe$_2$ (which is iso-structural to IrTe$_2$). The pre-edge intensity increases in going from FeTe, to IrTe$_2$, and eventually to AuTe$_2$, contrary to the expectation that the the number of absorption channels -- and thus the XAS intensity -- should be larger for lower $d$-shell occupation \cite{occupation}. Here we argue that the growing intensity trend observed in Fig.\,2(b) reflects an increase in covalence between ligand and transition-metal ions. The degree of covalence -- dependent on the charge-transfer energy $\Delta$ -- is expected to become larger for later transition-metals and higher valences, consistent with the observed evolution of the Te-$M_{4,5}$ pre-edge structure. This is similar to the intensity evolution of the oxygen $K$-edge pre-peak structure in transition-metal oxides, which is proportional to the unoccupied density-of-states (DOS) of the coupled ligand-oxygen-2$p$ and transition-metal-$d$ orbitals.

Figure\,2(c) shows the Te-$M_5$ pre-edge spectra taken at 200 and 300\,K. Light polarization was set parallel to the (1,0,0) axis, however in general no polarization dependence of the XAS signal was observed. As evidenced by these results, the Te-site partial DOS around the Fermi level $E_F$ -- corresponding to a photon energy of $\sim 569.7$\,eV in XAS -- is suppressed below the structural transition temperature $T_s$. At the same time, the partial DOS from 1.0 to 1.5\,eV above $E_F$ (corresponding to $570.7-571.2$\,eV in XAS) increases below $T_s$. As for the partial DOS above 2.0\,eV (above 571.7\,eV in XAS), and associated with the Te states hybridized with the Ir-$e_g$ manifold, it does not show a pronounced temperature dependence. The spectral changes observed across the transition seem consistent with the result of band structure calculations and dynamical mean-field theory (DMFT) \cite{Pascut13,Toriyama13,Qian13}, as well as with recent angle-resolved photoemission spectroscopy (ARPES) and resonant inelastic x-ray scattering (RIXS) studies of IrTe$_2$ \cite{Qian13,Ootsuki13}. These results suggest a Rice-Scott saddle-point-driven CDW instability \cite{rice,liu,kiss} associated with a Te-5$p$ van Hove singularity at $E_F$, which in the low-temperature (LT) phase is removed from $E_F$ due to the reconstruction of the electronic structure. The present XAS results for the unoccupied DOS are partly consistent with the ARPES/RIXS observations. However, the drastic change in the unoccupied DOS -- taking place up to 1.5\,eV above $E_F$ -- suggests that the simple saddle-point-driven CDW scenario is insufficient to fully describe the phase transition.

Next, we discuss the superstructure peak observed in REXS at $Q\!=\!(1/5,0,-1/5)$ in the LT phase. Fig.\,3(a) shows a $(H,0,-L)$ momentum scan  through the resonant peak at 200\,K and at a photon energy of 571.3\,eV, corresponding to the Te-$M_5$ pre-peak position. The signal is resonantly enhanced in the XAS pre-edge region, as evidenced by the REXS photon-energy dependence shown in Fig.\,3(c) and (d), indicating the active role of the covalent Te 5$p$\,--Ir 5$d$ bond-density in the CDW formation (the dip features found before the $M_{4,5}$ pre-edge structures will be analyzed in more detail in the discussion of Fig.\,4). As for the XAS main-edge region, x-ray absorption fine structure (XAFS) oscillatory behavior is observed, likely originating from local scattering of photoelectrons; however, the main-edge scattering intensity lacks a resonant character, which indicates that the $s$-$d$-$f$ hybrid band manifold does not participate in the ordering mechanism. 
\begin{figure}[b!]
\includegraphics[width=1\linewidth]{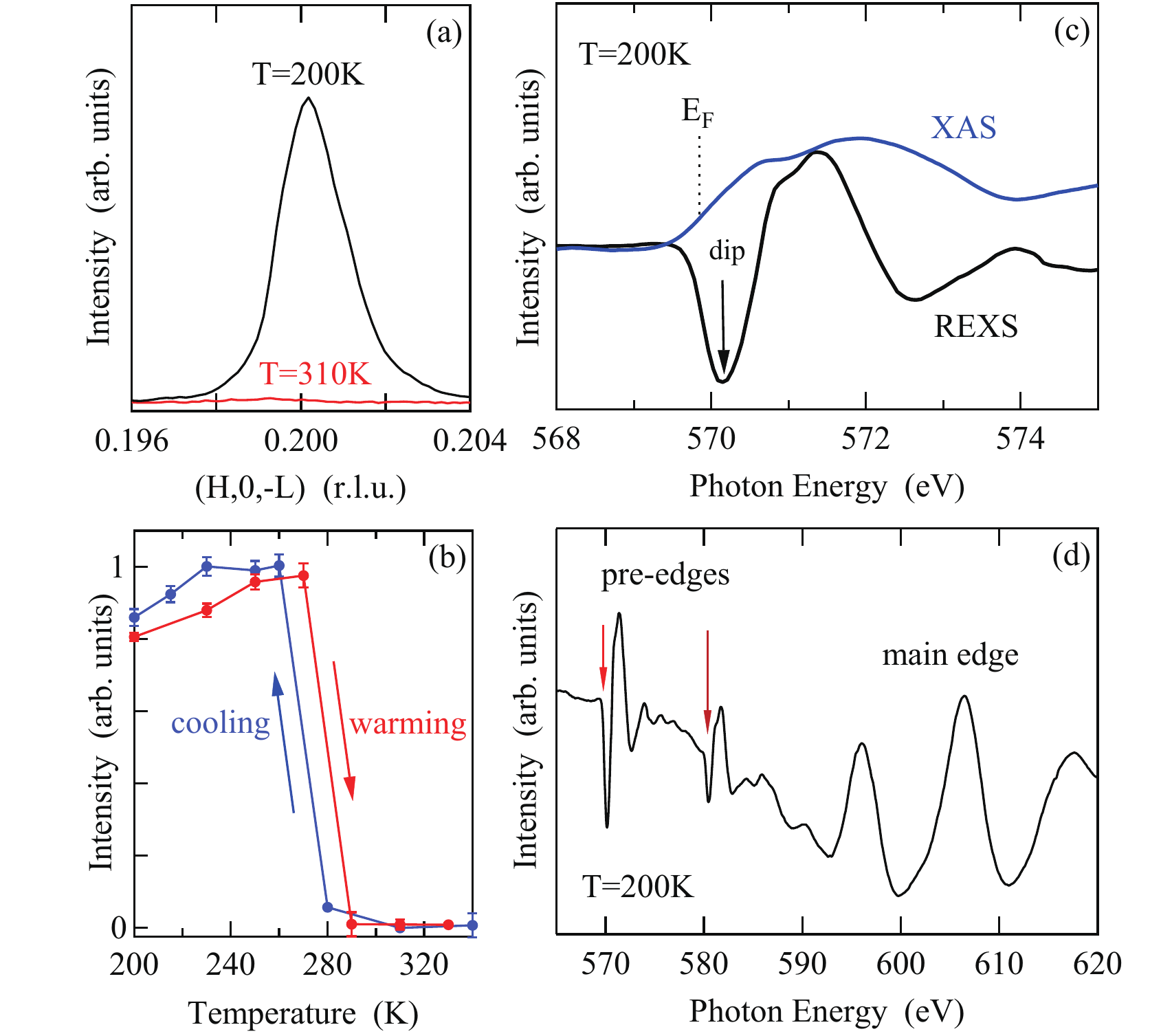}%
\caption{(color online). (a) REXS $(H,0,-L)$ scan through the $Q\!=\!(1/5,0,-1/5)$ superlattice peak measured on IrTe$_2$ at 200 and 310\,K, with 571.3\,eV photons. (b) Corresponding temperature dependence of the REXS intensity. (c) Comparison between REXS and XAS spectra in the $M_5$ pre-edge region at 200\,K; the arrow marks the dip structure before the REXS enhancement. (d) REXS spectrum in the entire energy range of the Te-$M$ edge x-ray absorption at 200\,K.}
\end{figure}

Figure\,3(b) shows the detailed temperature dependence of the $Q\!=\!(1/5,0,-1/5)$ superstructure peak amplitude in REXS, measured across $T_s$ during both cooling and warming cycles. The signal shows a sharp onset at $T_s$, consistent with the first-order character of the phase transition at $\sim\!280$\,K. In addition, a clear hysteretic behavior is also observed (the presence of a hysteretic behavior in XAS is discussed in the Supplemental Material \cite{supp}). This points to the formation of a multi-domain structure, where the CDW distortion  -- and in particular the shortening of one of the sides of the equilateral triangles forming the Ir sublattice in the $a-b$ plane -- may occur along any of the three triangular axes. The matching REXS intensity observed for the `slow' cooling and warming cycles in Fig.\,3(b), and conversely the mismatch and complex time and temperature evolution observed for `fast' cooling runs (see Supplemental Material \cite{supp}), suggest the presence of a `glassy' domain evolution that can reach equilibrium between the three possible domain orientations only during slow temperature cycles \cite{rexs}. This scenario is confirmed by LEED measurements on the very same sample which show -- along all three axes defining the triangular Ir sublattice -- analogous ($h$/5,0,-L) superstructure reflections at 200\,K [Fig.\,1(c)], but not at 300\,K [Fig.\,1(d)]. This domain structure, and its complex glassy evolution, might explain the controversy in the determination of the LT-phase structure \cite{Yang12,Cao13,Pascut13,Toriyama13}. 

Finally, we discuss the energy-dependent REXS lineshape shown in Fig.\,3(c) and (d). One should note that $E_F$ at $\sim 569.7$\,eV is located below the dip structure, while the resonant enhancement is maximum around 1\,eV above $E_F$. Therefore, the partial Te-DOS at $E_F$ contributes only weakly to the resonant enhancement seen in REXS, which instead mainly arises from the modulation of the unoccupied DOS around 1\,eV for the five structurally inequivalent sites [Fig.\,1(a)]. This result again challenges the conventional Fermi surface nesting picture as well as a van Hove singularity scenario, and instead agrees well with the results of band structure and DMFT calculations for the LT phase \cite{Pascut13,Toriyama13}. In particular in the calculation by Toriyama {\it et al.} \cite{Toriyama13}, the partial DOS of the Te(1)-$p_z$ orbital, which is hybridized with the dimerized Ir(3)-Ir(3) states, has indeed a sharp structure at around $\sim$1 eV; conversely, the DOS of Te(1)-$p_{x,y}$ and of all other Te-sites' $p$ orbitals is suppressed in this region. An electronic modulation involving the Te 5$p$ unoccupied DOS, coupled with the Ir site $t_{2g}$\!-orbital-order, is the best candidate to explain the REXS results in the LT phase.

For the quantitative analysis of the REXS lineshape, we use a methodology similar to the one introduced for the case of stripe-order in cuprates \cite{Achkar13}. The model relies on XAS measurements to determine the form factor $f(\omega)$ for the different Te sites (whereby any site-independent contribution will cancel out in REXS). The wavevector (\textbf{Q}) and photon-energy ($\omega$) dependent structure factor $S (\mathbf{Q},\omega)$ is subsequently constructed based on the spatial modulation of $f(\omega)$ at the different atomic positions $\mathbf{r}_j$:
\begin{equation} 
S (\mathbf{Q},\omega) = \sum_{j}{f_j(\omega) e^{-i \mathbf{Q} \cdot \mathbf{r}_j}},
\end{equation}
\begin{figure}[t!]
\includegraphics[width=1\linewidth]{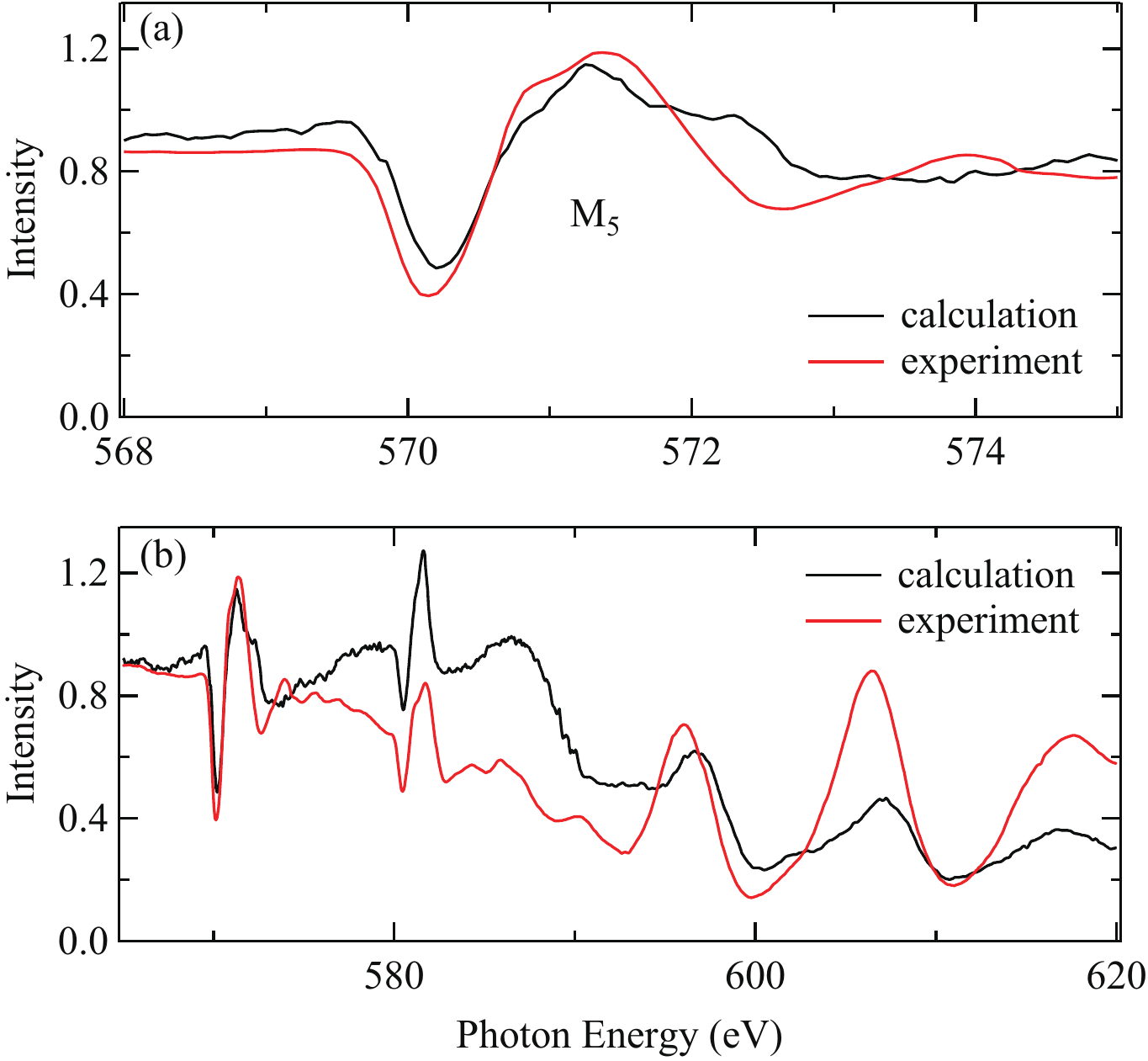}%
\caption{(color online). Calculated REXS intensity for the combination of a valence-modulation model (resonant term) with non-resonant lattice displacements, shown for: (a) the $M_5$ pre-edge region; and (b) the extended spectrum.}
\end{figure}
The experimental result is compared to three model calculations, where the major contribution to $S (\mathbf{Q},\omega)$ comes from, respectively: (i) lattice displacements, $\mathbf{r}_j\!=\!\mathbf{r}_j^0 + \delta \mathbf{r}_j$, where small displacements
are used for the Te and Ir lattice sites in the supermodulated structure; (ii) energy shifts, $f_j(\omega)\!=\!f(\omega + \delta \omega_j)$, where $\delta \omega_j$ is the spatial modulation of the energy of the Te-5$p$ state; and (iii) valence modulations, $f_j(\omega)\!=\! f(\omega,p+\delta p_j)$, where $\delta p_j$ is the variation in the local valence of the Te ions (further details on the three model calculations are given in the Supplemental Material \cite{supp}). The best agreement for the sharp dip features on the pre-edges, as well as the high-energy oscillatory behavior, is obtained using the valence (local DOS) modulation model, involving the covalent bonds between Te and Ir in the outermost shells. The comparison with experimental data is shown in Fig.\,4. Here, the form factors $f(\omega, p+\delta p_j)$ are assumed to modulate spatially as illustrated in the lower part of Fig.\,1. Proper atomic displacements -- contributing to the non-resonant terms -- are also embedded in the structure factor calculations. The present valence-modulation model reflects the periodic modulation of the Te 5$p$ orbitals coupled with the charge and/or orbital order at the Ir sites [Fig.\,1(a) and (b)], similar to the case of stripe-order in cuprates \cite{Achkar13,Fink09,Fink11,Abbamonte05,Hawthorn11}. Furthering this similarity, the IrTe$_2$ doping-pressure phase diagram exhibits a competitive interplay between superconductivity and other ordered phases \cite{Yang12,Pyon12,Fang13,Kamitani13,Kudo13,Kiswandhi13}; in analogy with recent studies of underdoped high-$T_c$ cuprates \cite{Ghiringhelli12,Comin13,Comin14,Comin15}, the role of stripe order as a candidate competing phase to superconductivity in IrTe$_2$ may also be probed -- across the superconducting transition -- by means of REXS at the Te 3$d$ $\rightarrow$ 5$p$ resonance. 

In conclusion, we have studied the ligand electronic states of IrTe$_2$ by XAS and REXS at the Te-${M}_{4,5}$ edge. The distinct pre-edge structure at the Te-$M_{4,5}$ edge in XAS reveals the prominent covalent Te 5$p$-Ir 5$d$ character of the near $E_F$ electronic structure (with ligand holes on the Te 5$p$ orbitals). A clear enhancement of REXS intensity at the $Q\!=\!(1/5,0,-1/5)$ superlattice reflection is observed below $T_s\!\sim\!280$\,K. We find the spatial modulation of the unoccupied DOS at Te sites -- covalently bonded to the Ir $t_{2g}$-orbitals --  to be responsible for the dominant contribution to the REXS intensity and, ultimately, for the stripe-like ordering formation in IrTe$_2$.

We thank M. Kobayashi for valuable discussions. This work was supported by the Max Planck - UBC Centre for Quantum Materials, the Killam, Alfred P. Sloan, Alexander von Humboldt, and NSERC's Steacie Memorial Fellowship Programs (A.D.), the Canada Research Chairs Program (A.D. and G.A.S.), NSERC, CFI, and CIFAR Quantum Materials. Part of the research described in this paper was performed at the Canadian Light Source, which is funded by the CFI, NSERC, NRC, CIHR, the Government of Saskatchewan, WD Canada, and the University of Saskatchewan.

\clearpage


\section*{Supplementary material}

# Supplemental Material

# Bond-Order and the Role of Ligand States in Stripe-Modulated IrTe$_2$

K. Takubo,[1,2,3] R. Comin,[1] D. Ootsuki,[4] T. Mizokawa,[4] H. Wadati,[5] Y. Takahashi,[6]
G. Shibata,[6] A. Fujimori,[6] R. Sutarto,[7] F. He,[7] S. Pyon,[8] K. Kudo,[8] M. Nohara,[8]
G. Levy,[1,2] I. Elfimov,[1,2] G. A. Sawatzky,[1,2] and A. Damascelli[1,2]

[1]Department of Physics & Astronomy, University of British Columbia, Vancouver, British Columbia V6T 1Z1, Canada
[2]Quantum Matter Institute, University of British Columbia, Vancouver, British Columbia V6T 1Z4, Canada
[3]Max Planck Institute for Solid State Research, Heisenbergstrasse 1, D-70569 Stuttgart, Germany
[4]Department of Physics & Department of Complexity Science and Engineering,
University of Tokyo, 5-1-5 Kashiwanoha, Chiba 277-8581, Japan
[5]Institute for Solid State Physics, University of Tokyo, Kashiwanoha 5-1-5, Chiba 277-8581, Japan
[6]Department of Physics, University of Tokyo, Hongo, Tokyo 113-0033, Japan
[7]Canadian Light Source, University of Saskatchewan, Saskatoon, Saskatchewan S7N 2V3 Canada
[8]Department of Physics, Okayama University, Okayama 700-8530, Japan

## I. METHODS

Resonant elastic x-ray scattering (REXS) and x-ray absorption spectroscopy (XAS) measurements were performed at the REIXS beamline of Canadian Light Source [1]. Single crystal samples of IrTe$_2$ were prepared using a self-flux method [2, 3]. The cleaved (001) plane was oriented at $\sim$54 degrees from the scattering plane in order to perform REXS measurements in the $Q_h = 2\pi(\frac{h}{a_H}, 0, \frac{h}{c_H})$ plane. Here, the reciprocal space indices $(h, k, l)$ refer to the high temperature trigonal (HTT) unit cell. For REXS measurements, the incident light was polarized along the [1, 0, -1] direction. IrTe$_2$ single crystals were cleaved in vacuum to minimize surface contamination effects; subsequently, the sample orientation was confirmed by detection of (1, 0, -1) Bragg reflections at 2.5 keV. Temperature dependence of normal-incidence XAS at the Te-$M_{4,5}$ edges was recorded both in the total electron yield (TEY) and total fluorescence yield (TFY) detection modes. XAS results using TEY mode showed no noticeable difference with respect to spectra acquired in TFY mode. XAS spectra – which were used to determine the form factor Im$\{f_j(\hbar\omega)\}$ – have been acquired in parallel with REXS scans and subsequently offset and scaled to calculated values of the absorption coefficient $\mu(\omega)$ (from NIST [4]) at the pre-edge and main-edge, in order to express $\mu(\omega)$ in units of $\mu m^{-1}$. Via the optical theorem, Im$\{f_j(\omega)\}$ is linearly proportional to the absorption coefficient $\mu(\omega)$, and Re$\{f_j(\omega)\}$ can be determined from Im$\{f_j(\omega)\}$ using Kramers-Kronig transformations. Accordingly, in order to express $f(\omega)$ in electrons/atom [Fig. S2(a)], experimental XAS spectra have been scaled and extrapolated to high and low energy using tabulated calculations of Im$\{f_j(\omega)\}$ above and below the absorption edge. Low-energy electron diffraction (LEED) measurements were performed on the same sample at UBC by using SPECS ErLEED 100. The momentum resolution was set to 0.02 $\text{Å}^{-1}$ with a low electron energy of 80 eV, where the signal intensity is maximized.

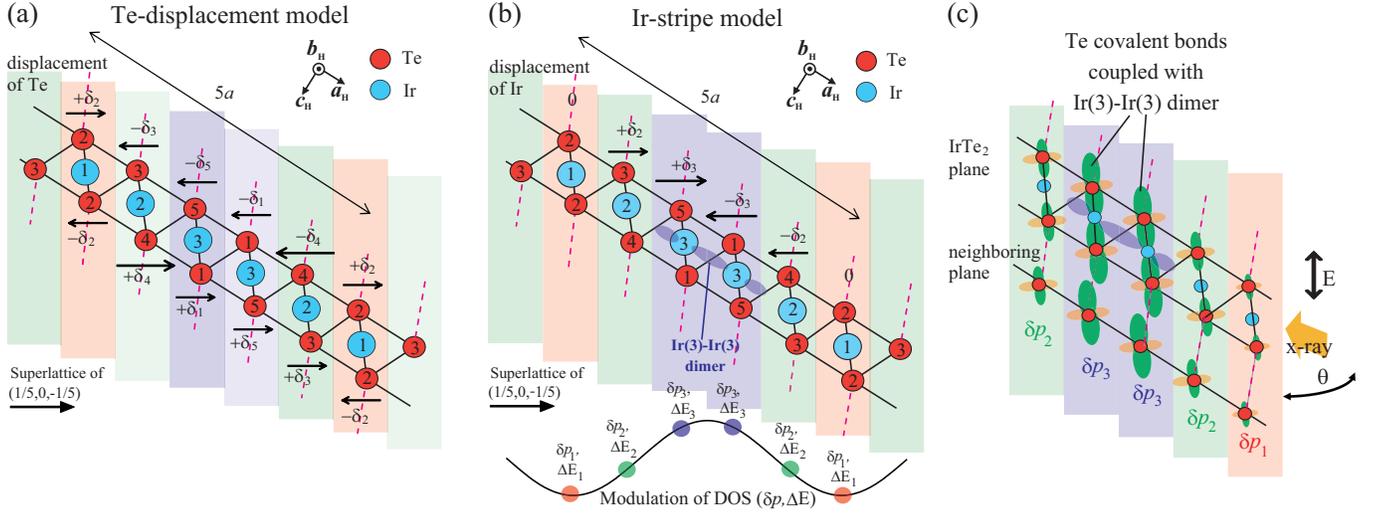

FIG. S1. Stripe-like ordering models with (1/5, 0, -1/5) wavevector for the Te and Ir sites. (a),(b) Schematic pictures of the (a) Te-displacement and (b) Ir-stripe models, with circled numeric labels denoting the inequivalent atoms. Arrows mark the displacements along [1,0,-1] for the Te and Ir sites. The modulation of the density of state estimated from a DMFT calculation [6] is shown at the bottom, and the dimerization between Ir(3)-Ir(3) sites is also indicated. (c) Experimental geometry of REXS and covalent bonds on the Te sites hybridized with the Ir sites. Because of the experimental geometry and incident polarization, the scattering experiment is particular sensitive to the green Te-Te bonds between the IrTe$_2$ planes.

## II. CALCULATION OF REXS INTENSITY

The calculation of the REXS intensity is structured similarly to the method successfully applied to stripe-ordered cuprates in Ref. 5. The calculation is performed for three different options, namely: (1) *lattice displacement model*, where small displacement are used for the Te and Ir lattice sites in the supermodulated structure; (2) *energy shift model*, assuming a spatial modulation in the energy of the Te 5$p$ state; and (3) *valence modulation model*, corresponding to a periodic variation in the local valence of Te ions. These models are subsequently implemented in the calculation of the structure factor $S(\mathbf{Q},\omega)$, which is written generally as:

$$S(\mathbf{Q},\omega) = \sum_j f_j(\omega) e^{-i\mathbf{Q}\cdot(\mathbf{r}_j^0 + \delta\mathbf{r}_j)}, \qquad (1)$$

where $f_j$ is the atomic form factor at the lattice position $j$, $\omega$ is the photon energy, $\mathbf{Q}$ is the scattering vector, $r_j^0$ is the position vector in the undistorted structure at site $j$ and $\delta r_j$ is the displacement from the lattice position due to the structural modulation. The atomic form factor can also depend on additional parameters related to the electronic structure of the atom at $j$, such as the local charge density or energy levels; these factors are explicitly included in the respective models. More specifically, all the energy dependent terms are included in the atomic form factor $f_j(\omega)$, while the atomic positions or displacements are of course energy-independent.

In order to determine the structure factor for these various models, we rest on a few assumptions: (i) we take $Q_H = 2\pi(\frac{1}{5a_H}, 0, -\frac{1}{5c_H})$ in the HTT phase, corresponding to $Q_L = 2\pi(0, 0, \frac{1}{c_L})$ in the low temperature triclinic (LTT) structure; (ii) the scattering measurements at the Te $M_{4,5}$ pre-edges are sensitive to the covalent states formed by the Te 5$p$ orbitals hybridized with the Ir 5$d$ orbitals [see Fig. S1(c)]; (iii) we assume charge scattering to be dominant and therefore neglect the off diagonal matrix elements of the form factor. Assumption (iii) is validated by the fact that the scattering intensity becomes very weak for $\pi$-polarized incident x-rays, amounting to $\sim 1/35$ of that for $\sigma$ polarization, which is comparable to the factor of $\cos^2(2\theta) \sim 1/40$ for the charge scattering at a detector angle of $2\theta \sim 99$ degrees. For what concerns the structure, in the LTT unit cell there are 3 inequivalent Ir sites forming a sequence -Ir(1)-Ir(2)-Ir(3)-Ir(3)-Ir(2)- perpendicular to the stripe direction in the IrTe$_2$ plane [Fig. S1(a) and Table I] [6, 7]. Here parentheses $(j)$ denote the site number. In addition, there are 5 inequivalent Te atoms in the unit cell, with a sequence -Te(2)-Te(3)-Te(5)-Te(1)-Te(4)- again perpendicular to the stripe direction. When the form factors $f_j(\omega)$ depend on the inequivalent Te sites, we use the denomination of 'Te-displacement model'. However, the form factors can in principle also depend on the coordinated Ir sites, since these states are hybridized with Te states and are therefore involved in the stripe modulation. In this last case, we speak of a 'Ir-stripe model' [see also Fig. S1(b)].

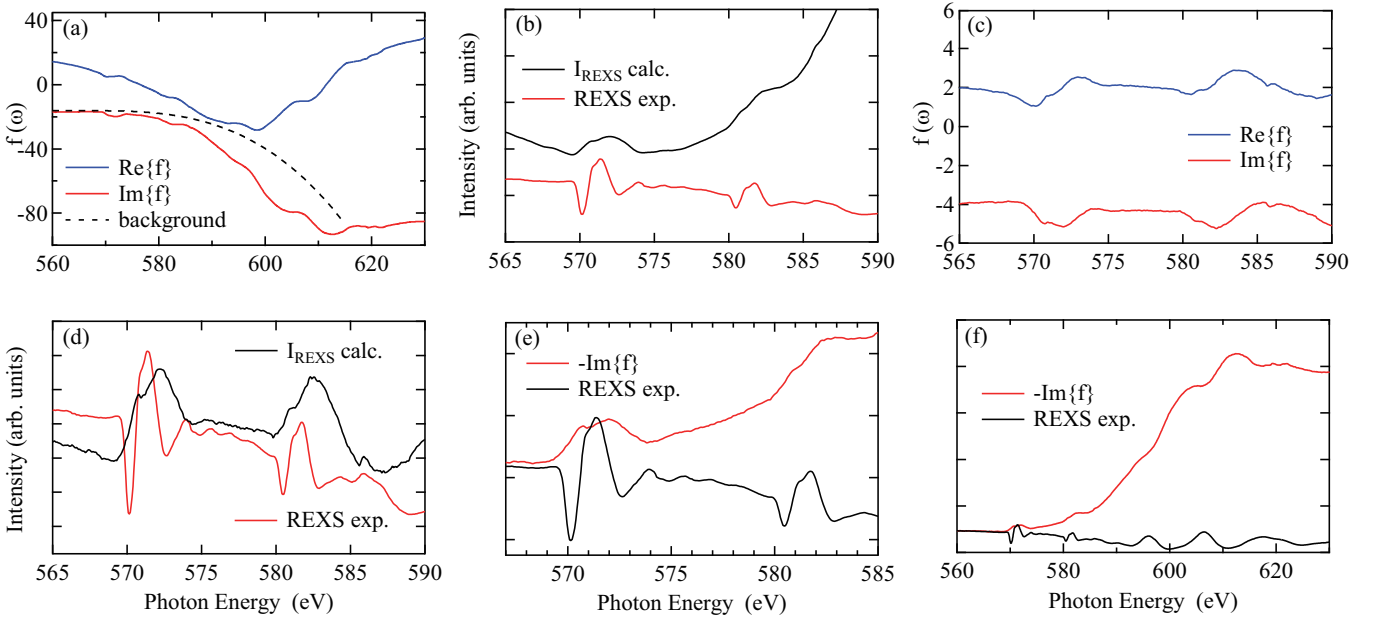

FIG. S2. Form factors $f(\omega)$ and the calculated REXS intensity using the lattice displacement model. (a) Real (blue) and imaginary (red) parts of $f(\omega)$. The black dashed line denotes the background $\sim a\omega^3$ for the main edge. (b) Calculated $I_{\rm REXS}$ (black) and experimental REXS (red) spectra. (c) Real (blue) and imaginary (red) parts of $f(\omega)$ after background subtraction. (d) $I_s$ as calculated from the background subtracted $f(\omega)$. (e),(f) Comparison of the energy dependence of -Im$f$ (red) and REXS (black) in the (e) pre-edges region and (f) entire regions.

**Lattice displacement model**

For the lattice displacement model, $f_j$ is the same at each site, but lattice positions are displaced, i.e. $\boldsymbol{r}_j = \boldsymbol{r}_j^0 + \delta \boldsymbol{r}_j$. Considering a chain of 5 Te sites separated by $(a_H, 0, -c_H) \sim (0, 0, c_L/5)$, the structure factor is given by:

$$S = f(\omega)\left[e^{\frac{-2i\pi\delta_2}{c_L}} + e^{\frac{-2i\pi(c_L/5+\delta_3)}{c_L}} + e^{\frac{-2i\pi(2c_L/5+\delta_5)}{c_L}} + e^{\frac{-2i\pi(3c_L/5+\delta_1)}{c_L}} + e^{\frac{-2i\pi(4c_L/5+\delta_4)}{c_L}}\right]. \quad (2)$$

In the limit of small displacements, we can expand the exponential terms to first order and write:

$$S = f(\omega)\frac{-2i\pi}{c_L}\left(\delta_2 + e^{\frac{-2i\pi}{5}}\delta_3 + e^{\frac{-4i\pi}{5}}\delta_5 + e^{\frac{-6i\pi}{5}}\delta_1 + e^{\frac{-8i\pi}{5}}\delta_4\right), \quad (3)$$

which gives a scattering intensity that can be expressed as:

$$I_{\rm REXS} = \frac{C|S|^2}{\mu(\omega)} \cong \frac{C}{\mu(\omega)}\frac{4\pi^2}{c_L^2}|f(\omega)|^2\left|\delta_2 + e^{\frac{-2i\pi}{5}}\delta_3 + e^{\frac{-4i\pi}{5}}\delta_5 + e^{\frac{-6i\pi}{5}}\delta_1 + e^{\frac{-8i\pi}{5}}\delta_4\right|^2 \propto \frac{|f(\omega)|^2}{\mu(\omega)}. \quad (4)$$

This result holds even if one includes higher order terms in the series expansion. Moreover, the magnitude of the displacements has no impact on the energy dependence of the calculated scattering intensity. Although we have depicted the Te-displacement model in Fig. S1(a), the generality of this result implies that the same energy dependence in $I_S$ is obtained for both the Te-displacement and Ir-stripe models.

Figure S2(c) shows the result from the simulation of the energy-dependent REXS intensity using the lattice displacement model, i.e. $I_S(\omega) \propto |f(\omega)|^2/\mu(\omega)$. Clearly, the lineshape of the pre-edge region is significantly different from the experimental result [see Fig. S2(b)]. This discrepancy is partly due to the tail of the strong main edge in Re$\{f(\omega)\}$ derived from the K-K transformation, which extends down to the pre-edge region. Here, the final states of the main-edge are the $sdf$ hybrid states at the Te sites. Therefore, the main edge should only contribute very weakly to the resonance enhancement in the present REXS experiment [see also Fig. S2(f)]. To get rid of this contribution, a proper background of the form $a\omega^3$ (for $\hbar\omega >$570eV) was subtracted from Im$\{f(\omega)\}$ [Fig. S2(d)] [8]. Although the REXS signal is roughly reproduced after performing the background subtraction, the sharp dip features located around $\sim 570$ eV and $\sim 580$ eV in the pre-edges remain poorly matched. The predominant reason for this discrepancy resides on the broad peak structures with $\sim$3 eV width in Im$\{f(\omega)\}$ as derived from XAS, which are much broader

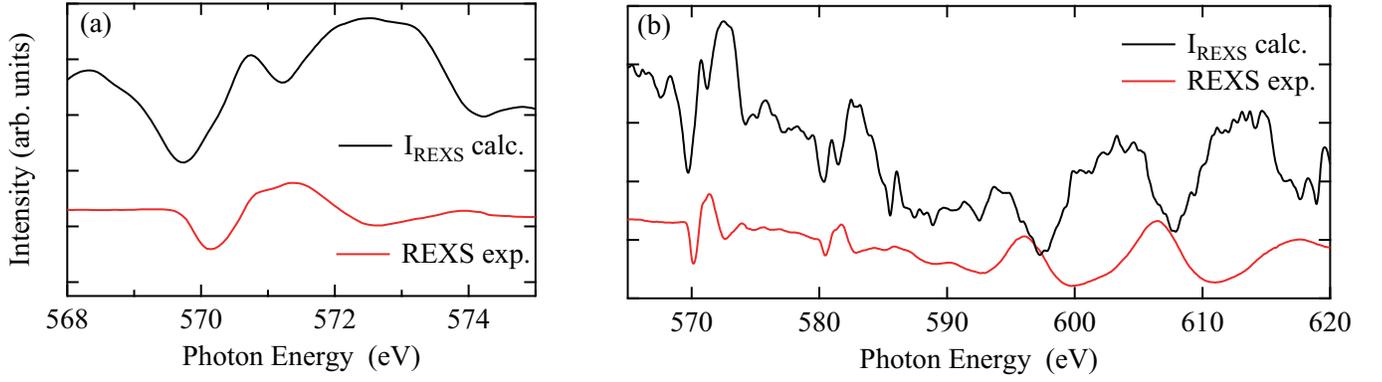

FIG. S3. Calculated (black) and experimental (red) REXS intensity using the energy shift model at (a) the $M_5$ pre-edge and for (b) the extended spectrum.

than those in REXS [see Fig. S2(e)], a phenomenon which has been observed in other systems [9, 10]. As discussed in the main text, however, this broad feature with 3-4 eV peak-width in XAS directly reflects the unoccupied $5p$ density of state on the Te sites and is consistent with the band structure calculations in Ref. 6 and 7. Therefore, this result suggests that the lattice displacement model cannot be reconciled under any assumption with the energy-dependent lineshape seen in REXS. The modulation of the electronic structure at the Te sites should therefore be taken into the consideration, as will be done below.

**Energy shift model**
The energy shift model takes into account spatial variations in transition energies at the Te sites. In the case of the Ir-stripe model [Fig. S1(b)], there are three inequivalent sites with different transition energies, $\hbar\omega_1 = \hbar\omega + \Delta E_1$, $\hbar\omega_2 = \hbar\omega + \Delta E_2$, and $\hbar\omega_3 = \hbar\omega + \Delta E_3$. The structure factor is thus given by:

$$S = f(\hbar\omega + \Delta E_1) + e^{\frac{-2i\pi}{5}} f(\hbar\omega + \Delta E_2) + e^{\frac{-4i\pi}{5}} f(\hbar\omega + \Delta E_3) + e^{\frac{-6i\pi}{5}} f(\hbar\omega + \Delta E_3) + e^{\frac{-8i\pi}{5}} f(\hbar\omega + \Delta E_2)$$
$$= f(\hbar\omega + \Delta E_1) + 2\cos\left(\frac{-2\pi}{5}\right) f(\hbar\omega + \Delta E_2) + 2\cos\left(\frac{-4\pi}{5}\right) f(\hbar\omega + \Delta E_3), \quad (5)$$

and the scattering intensity can be written as:

$$I_{\text{REXS}} = \frac{C}{\mu(\omega)} |f(\hbar\omega + \Delta E_1) + 2\cos\left(\frac{-2\pi}{5}\right) f(\hbar\omega + \Delta E_2) + 2\cos\left(\frac{-4\pi}{5}\right) f(\hbar\omega + \Delta E_3)|^2. \quad (6)$$

In the calculation shown in Fig. S3(a), we set $\Delta E_1 = -0.3$ eV, $\Delta E_2 = -0.05$ eV, and $\Delta E_3 = 0.2$ eV, i.e. the energy shifts are assumed to be proportional to the density of states (DOS) modulation as given in Refs. 6 and 7. The DOS at $E_F$ and the energy shifts $\Delta E_j$ perpendicular to the stripes are indeed assumed to follow a periodic modulation, shown in the lower part of Fig. S1(b). In the calculation, any value $0 < |\Delta E_j| \leq 0.5$ eV with a sine-like modulation produces a similar energy dependence in $I_{\text{REXS}}$, while otherwise only changing the overall amplitude of the calculated profiles. The values of $\Delta E_j$ affect the magnitude, which scales as $\Delta E_j$, but not the energy-dependent line shape of the calculated intensity [5]. The results of the calculation are plotted in Fig. S3. The sharp dips before the pre-edge region are here better reproduced as compared to the lattice displacement model. However, the phase of the high energy oscillatory structure does not match the experimental one. Moreover, the line shape around 573 eV is not well reproduced, whereby another dip structure is observed on REXS with an energy corresponding to the $e_g$ state seen in XAS. The modulation of the $e_g$ state (together with the high energy oscillatory structures) would require opposite energy shifts, and therefore cannot be described within this model.

In case of the Te-displacement model, the structural factor and scattering intensity are instead given by:

$$S = [f(\hbar\omega + \Delta E_2) + e^{\frac{-2i\pi}{5}} f(\hbar\omega + \Delta E_3) + e^{\frac{-4i\pi}{5}} f(\hbar\omega + \Delta E_5) + e^{\frac{-6i\pi}{5}} f(\hbar\omega + \Delta E_1) + e^{\frac{-8i\pi}{5}} f(\hbar\omega + \Delta E_4)], \quad (7)$$

$$I_{\text{REXS}} = \frac{C}{\mu(\omega)} |f_2 + e^{\frac{-2i\pi}{5}} f_3 + e^{\frac{-4i\pi}{5}} f_5 + e^{\frac{-6i\pi}{5}} f_1 + e^{\frac{-8i\pi}{5}} f_4|^2. \quad (8)$$

However, the magnitudes of the energy shifts for the Te atoms are hard to estimate, since there are 5 inequivalent atoms and therefore more fitting parameters. Here, the result is omitted.

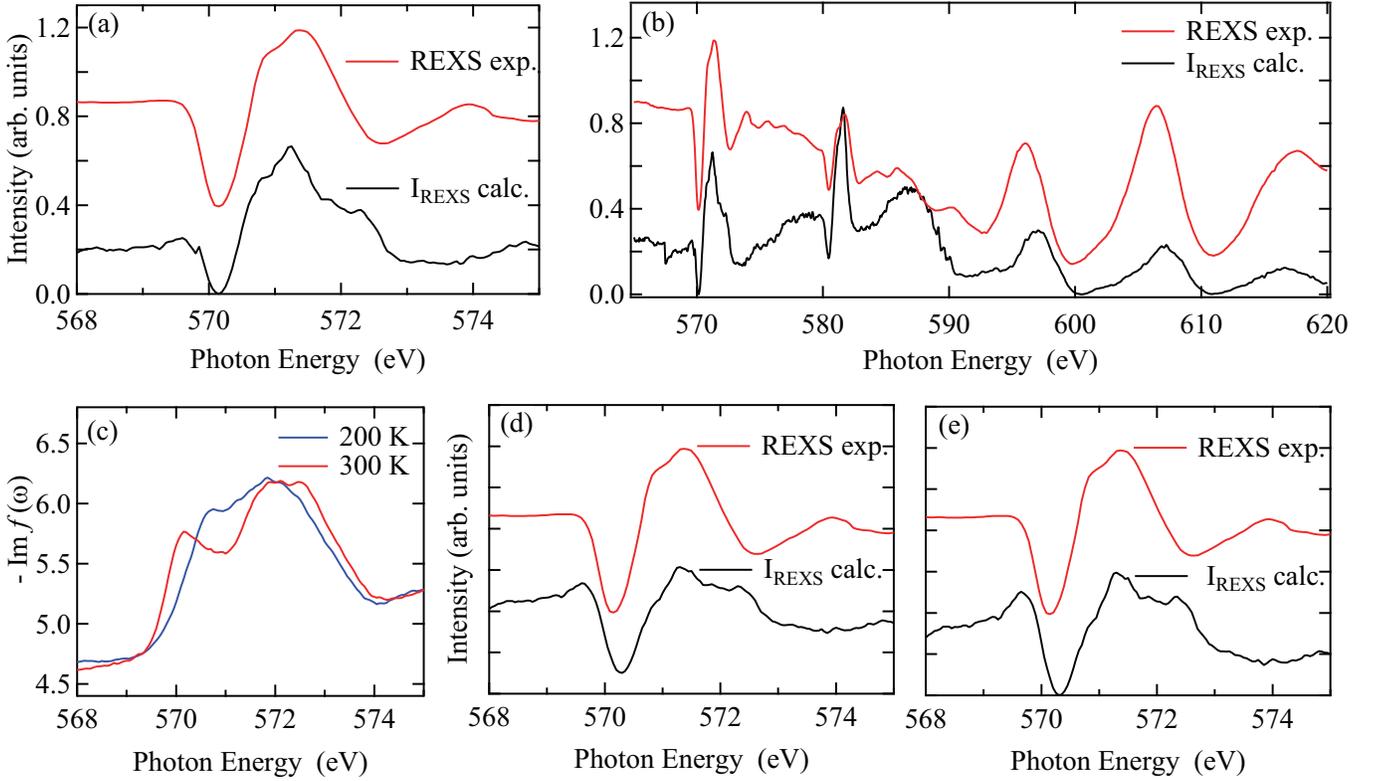

FIG. S4. Comparison between experimental results and REXS intensity calculated within the Ir-stripe valence modulation model at (a) the $M_5$ pre-edge and for (b) the full spectrum, with parameters $\delta p_1$ = -0.9, $\delta p_2$ = -0.15, and $\delta p_3$ =0.6. (c) $-\mathrm{Im} f(\omega)$ at 200 and 300 K. (d),(e) Comparison between experimental and calculated intensity at the $M_5$ pre-edge within the Te-displacement valence modulation model for the following parameters: (d) $\delta p_2$ =-0.9, $\delta p_3$ = 0.1, $\delta p_5$ = 0.1, $\delta p_1$ =0.1; and $\delta p_4$ =0.6 (proportional to the Te displacement along the $z$ axis); (e) $\delta p_2$ =-0.6, $\delta p_3$ = -0.6, $\delta p_5$ = -0.15, $\delta p_1$ =0.9, and $\delta p_4$ =-0.15 (for modulation centered at the Te(1) site coupled with the Ir-Ir dimerization, but against the Ir-stripe formation).

**Valence modulation model**

Finally, the valence (or local DOS) modulation model is considered. The difference with the energy shift model is that here we use the spatial modulation of the local DOS in place of the energy shifts; in this case, the structure factor for the Ir-stripe model is given by:

$$S = f(\omega, p+\delta p_1) + e^{\frac{-2i\pi}{5}} f(\omega, p+\delta p_2) + e^{\frac{-4i\pi}{5}} f(\omega, p+\delta p_3) + e^{\frac{-6i\pi}{5}} f(\omega, p+\delta p_3) + e^{\frac{-8i\pi}{5}} f(\omega, p+\delta p_2)$$
$$= f(\omega, p+\delta p_1) + 2\cos\left(\frac{-2\pi}{5}\right) f(\omega, p+\delta p_2) + 2\cos\left(\frac{-4\pi}{5}\right) f(\omega, p+\delta p_3), \qquad (9)$$

while for the Te-displacement model we have:

$$S = [f(\omega, p+\delta p_2) + e^{\frac{-2i\pi}{5}} f(\omega, p+\delta p_3) + e^{\frac{-4i\pi}{5}} f(\omega, p+\delta p_5) + e^{\frac{-6i\pi}{5}} f(\omega, p+\delta p_1) + e^{\frac{-8i\pi}{5}} f(\omega, p+\delta p_4)]. \qquad (10)$$

We determine $f(\omega, p_j)$ as a function of the local DOS modulation $p_j$ by performing a linear extrapolation from the $f(\omega)$ measured with XAS at 300 K (HT phase) and 200 K (LT phase)[see Fig. S4(c)]:

$$f(\omega, p_j) \cong f(\omega, T=300\mathrm{K}) + A_j[f(\omega, T=200\mathrm{K}) - f(\omega, T=300\mathrm{K})]. \qquad (11)$$

Here, $A_j$ is proportional to the DOS modulation $\delta p_j$ at site $j$. We assume that $f(\omega)$ at $T$=200 and 300 K corresponds to ordered and non-ordered phases, respectively, and the extrapolation in Eq. 11 is used to express the covalent densities (or local DOS), namely $f_j(\omega)$, at each Te sites. In addition, it should be noted that $p_j$ in the present calculation represents the spatial modulation of the entire local unoccupied DOS for the Te($j$) sites including the high energy structures, and not only of the local hole doping levels near $E_F$.

The scattering intensity for the Ir-stripe model is then given by:

$$I_s = \frac{C}{\mu(\omega)} |f(\omega, p_1) + 2\cos\left(\frac{-2\pi}{5}\right) f(\omega, p_2) + 2\cos\left(\frac{-4\pi}{5}\right) f(\omega, p_3)|^2, \qquad (12)$$

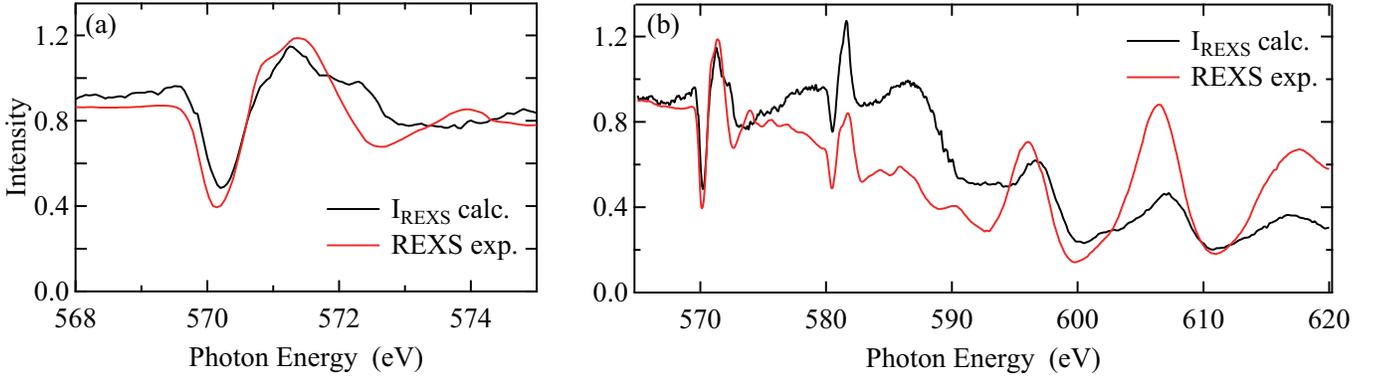

FIG. S5. Calculated (black) and experimental (red) REXS intensity for the sum of non-resonant lattice modulation and resonant valence modulation models at (a) the $M_5$ pre-edge and (b) for the extended spectrum.

whereas for the Te-displacement model we have:

$$I_s = \frac{C}{\mu(\omega)}|f(\omega,p_2) + e^{\frac{-2i\pi}{5}}f(\omega,p_3) + e^{\frac{-4i\pi}{5}}f(\omega,p_5) + e^{\frac{-6i\pi}{5}}f(\omega,p_1) + e^{\frac{-8i\pi}{5}}f(\omega,p_4)|^2. \quad (13)$$

The comparison between experimental results and REXS intensity calculated for both Ir-stripe and Te-displacement models is shown in Fig. S4. While the Ir-stripe model closely reproduces the sharp dips before and after the enhancement in the pre-edge region [Fig. S4(a)], and also matches the high energy oscillatory structure [Fig. S4(b)], for the Te-displacement model the agreement is not as satisfactory [e.g., the dip-energy exceeds the experimental value for both parameter sets, and the intensity variation from below to just above the dip feature is also not well reproduced, as shown in Fig. S4(d) and S4(e)]. In this context, one may ascribe the dip and hump structures in REXS to a simple strain wave [11], and assume that these structures do not bear any information about the charge and/or orbital order. However, the pre-edge structure in XAS represents the unoccupied density of states for the covalent bonds bridging between Ir and Te orbitals, or equivalently the ligand holes on the Te sites. Here we outline a few elements supporting the predominant role of covalence over that of Te atomic displacements: (i) the symmetry of these Te states in the outermost shells depends on the crystal field of the IrTe$_6$ octahedra, rather than the positions of each Te atoms; (ii) the lattice displacements $\delta_j$ for each Te atom are only a few percent of the lattice spacing (see Table I); (iii) Te displacements are not similar to those for the coordinated Ir sites, since the former are affected by the Jahn-Teller distortion and the interaction between the planes through the Te-Te bonds, as well as the stripe formation perpendicular to the [1/5,0,-1/5] direction – indeed the Te(5) atoms move in the negative direction along [1/5,0,-1/5] (corresponding to the $z$ axis in the LTT unit cell), while the coordinated Ir(3) atoms move in the positive direction due to the dimer formation between Ir(3)-Ir(3), as shown in Fig. S1 and Table I [6, 7].

In summary, the best agreement between calculated and experimental REXS intensity is obtained for the Ir-stripe valence modulation model. Merging all of these elements together, it is reasonable to assume that REXS energy dependence in the pre-edge region originates from local variations in the form factor $f(\omega)$ [see Fig. S4(c) for imaginary part], rather than from the lattice displacement of Eq. (1); our calculations qualitatively support this scenario.

| Atom | x(220K) | y(220K) | z(220K) | $\delta/c_L$(220K) | x(20K) | y(20K) | z(20K) | $\delta/c_L$(20K) |
|---|---|---|---|---|---|---|---|---|
| Ir(1) | 0 | 0 | 0 | 0 | 0 | 0 | 0 | 0 |
| Ir(2) | 0.42570 | 0.78722 | 0.20327 | ±0.00327 | 0.36069 | 0.21315 | 0.203394 | ±0.003394 |
| Ir(3) | 0.14156 | 0.42983 | 0.58879 | ±0.01121 | -0.29000 | 0.43031 | 0.411623 | ±0.011623 |
| Te(2) | 0.36780 | 0.72958 | 0.01680 | +0.01680 | 0.36127 | 0.27091 | 0.01679 | +0.01679 |
| Te(3) | 0.05400 | 0.05471 | 0.18480 | −0.01520 | -0.00026 | -0.05426 | 0.18477 | −0.01523 |
| Te(5) | 0.51730 | 0.15875 | 0.61221 | −0.01221 | 0.35701 | 0.15939 | 0.38782 | −0.01218 |
| Te(1) | 0.21740 | 0.30048 | 0.41130 | −0.01130 | 0.08129 | 0.70076 | 0.41086 | −0.01086 |
| Te(4) | 0.20320 | 0.47996 | 0.77765 | −0.02235 | -0.27774 | 0.48038 | 0.22267 | −0.02267 |

TABLE I. Atomic positions and displacements in lattice parameter units at 220 K [6] and 20 K [7]. The space group is $P$-$I$, the lattice parameters are $a$=3.9548(2) Å, $b$=6.6542(4) Å, and $c$=14.4345(7) Å at 220 K.

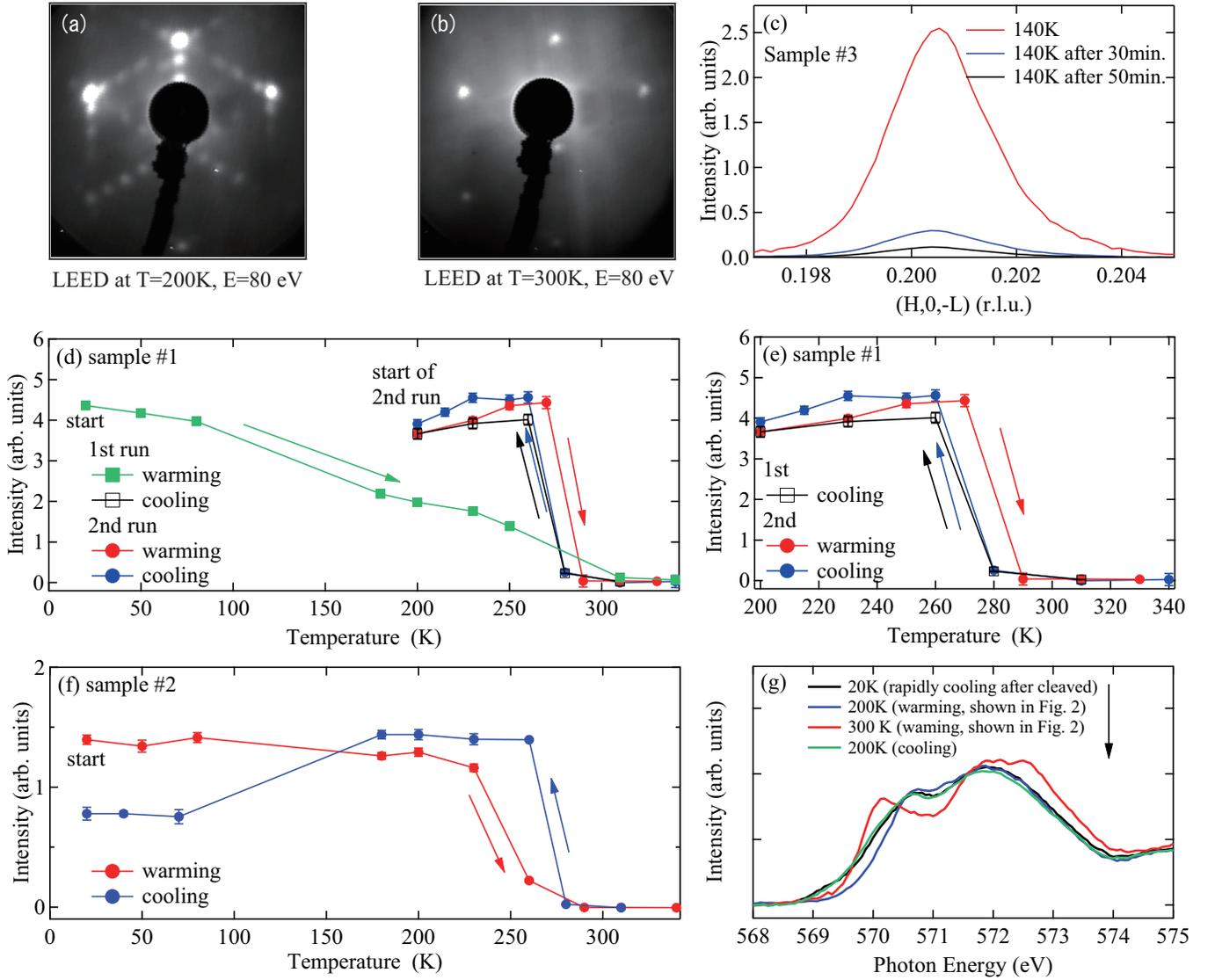

FIG. S6. Temperature dependence the superlattice peaks of IrTe$_2$. (a),(b) LEED patterns for IrTe$_2$ at (a) 200 K and (b) 300 K. (c) Profile through (0.2,0,-0.2) at $\hbar\omega$=571.3 eV and $T$=140 K (red); same scan, acquired after 30 (blue) and 50 minutes (black) at fixed temperature. (d)-(f) Temperature dependence of the REXS intensity of the (0.2,0,-0.2) reflection at $\hbar\omega$=571.3 eV for: (d) for sample #1 during the 1st and 2nd attempts; (e) the sample #1 during the 2nd attempts; and (f) the sample #2. Sample #2 was warmed up and cooled down as fast as possible between 80 K and 200 K. The 2nd attempts in (e) are also shown in the main text. (g) XAS spectra at various temperatures. The arrow denotes the measurement sequence.

**Sum of non-resonant lattice modulation and resonant valence modulation**

Although the energy dependent REXS signal in the Te 3$d$ pre-edge region is essentially reproduced by the valence modulation model, the non-resonant contribution, or baseline does not match comparably well. Since the superstructural order involves lattice distortions as well, non-resonant terms will also contribute to the baseline in the energy-dependent REXS profile (see Fig. S5). Both the resonant and non-resonant contribution can be estimated by combining the non-resonant lattice modulation and the resonant valence modulation of the electronic state. The total structure factor is then written as:

$$S = \sum_j f_c e^{-i\boldsymbol{Q}\cdot(\boldsymbol{r}_j+\delta\boldsymbol{r}_j)} + \sum_j f(\hbar\omega,p_j) e^{-i\boldsymbol{Q}\cdot\boldsymbol{r}_j}. \tag{14}$$

Here, $f_c$ is constant and assumed to be the form factor below 560 eV from the tabulated data of NIST [4]. The first term corresponds to the atomic scattering from the inner shell for Te and Ir atoms. On the other hand, the second term denotes the hybridized orbitals of the Te-Ir bonds in the outermost shells discussed in the previous section.

Then, the scattering intensity goes as:

$$I_S \cong \frac{C}{\mu(\omega)} \left| (\alpha\delta_2 f_c + f_2) + e^{\frac{-2i\pi}{5}}(\alpha\delta_3 f_c + f_3) + e^{\frac{-4i\pi}{5}}(\alpha\delta_5 f_c + f_5) + e^{\frac{-6i\pi}{5}}(\alpha\delta_1 f_c + f_1) + e^{\frac{-8i\pi}{5}}(\alpha\delta_4 f_c + f_4) \right|^2 \quad (15)$$

where $\alpha = -2i\pi/c_L$.

The result of the calculation is plotted in Fig. S5, and also in Fig. 4 of the main text. The energy dependent line shape is well-reproduced by this model, including the high energy XAFS-like oscillatory structures for the main edge. The ratio between the peak amplitudes and baseline is in close agreement with the experiment, and depends on $\delta p_j$. A least square fitting analysis returns reasonable values of $\delta p_2 = -0.9$, $\delta p_3 = \delta p_4 = -0.15$, and $\delta p_5 = \delta p_1 = 0.6$, in agreement with the periodic local DOS modulation for the Ir-stripe order [6].

## III. TEMPERATURE DEPENDENCE

Figure S6(a) and (b) show LEED images at 200 K and 300 K. The $(h/5,0,-L)$ superstructural satellites are observed at 200 K while are absent at 300 K. These features are detected along all three triangular sublattices on this sample, indicating a complex glassy evolution seemingly consistent with the REXS result.

As discussed in the main text, the temperature dependence of the (0.2,0,-0.2) superstructural reflection in REXS above 200 K indicates the clear first order character of the transition at $T_s$ in IrTe$_2$. The samples were cleaved at room temperature and rapidly cooled down to 20 K in order to see the (0.2,0,-0.2) reflection. However, the temperature dependence below 200 K shows a curios re-entrant behavior. When the temperature ranges between 80 K and 200 K, the REXS intensity decreases with time as shown in Fig. S6(c). This behavior is not observed below 80 K or above 200 K as shown in Fig. S6(d)-(f) and likely does not originate from radiation damage or contamination effect, since the scattering intensity during the cooling run below $T_s$ always recovers to the same value recorded during warming cycles. This behavior can possibly correlate to the development of additional modulations with a period of 1/8 and/or 1/11 below 200 K, as reported in a recent STM study [12]. Further REXS studies below 200 K could potentially confirm these additional modulations.

The temperature dependence of XAS also exhibits a large hysteresis, as shown in Fig. 6(g). The intensity below 570 eV, namely the DOS at $E_F$ corresponding to the structure around (or lower than) 569.7 eV, remains approximately constant in the spectra at 20 K (black line) and at 200 K during the cooling run (green line), while being substantially suppressed at 200 K during the warming run (blue line).

---




\begin{references}
\bibitem{occupation}Assuming a Te oxidation state of $-$2, the transition-metal valences are formally Fe$^{2+}$($3d^6$), Ir$^{4+}$($5d^5$), and Au$^{4+}$($5d^7$), which would suggest AuTe$_2$/FeTe/IrTe$_2$ as sequence of progressively increasing intensity purely based on the filling of the transition-metal $d$ orbitals.
\bibitem{rexs} We note that REXS and LEED, with their spot size of about 500\,$\mu$m and 2\,mm diameter, respectively, effectively average over the multi-domain mosaic. Also, due to geometrical constraints of the sample manipulator, in REXS it was not possible to directly probe the superstructure wavevector along different directions; the presence of domains can thus be inferred only from the intensity evolution of the $Q\!=\!(1/5,0,-1/5)$ reflection.
\end{references}
\end{document}